\documentclass[aps,prd,onecolumn,nofootinbib]{revtex4-2} %%,superscriptaddress  %%onecolumn oder twocolumn
\usepackage{titlesec}
\titleformat*{\section}{\LARGE\bfseries}
\titleformat*{\subsection}{\Large\bfseries}
\titleformat*{\subsubsection}{\large\bfseries}%\itshape

\usepackage{epsfig}
\usepackage{enumitem}

\usepackage{graphics}     
\usepackage{graphicx}     
\usepackage{subfigure}
\usepackage{longtable}     
\usepackage{url}          
\usepackage{bm}           
\usepackage{natbib}
\usepackage{textpos}
\usepackage{amsfonts,amsmath,amssymb,mathrsfs,bm,amsthm}
\usepackage{float}
\usepackage{booktabs}
\usepackage{array}
\usepackage{tabu}
\usepackage{dcolumn}
\usepackage{rotating}
\usepackage{ulem}

\newcommand{\RN}[1]{%
  \textup{\uppercase\expandafter{\romannumeral#1}}%
}

\usepackage{nicefrac}
\usepackage{amsthm}
\usepackage{color}
\usepackage{cancel}
\usepackage{appendix}
\usepackage{makecell}
\usepackage{upgreek}

\usepackage[colorlinks=true,linkcolor=black,citecolor=blue,urlcolor=blue,bookmarksopen]{hyperref}
\newcommand{\appropto}{\mathrel{\vcenter{
  \offinterlineskip\halign{\hfil$##$\cr
    \propto\cr\noalign{\kern2pt}\sim\cr\noalign{\kern-2pt}}}}}
\renewcommand{\v}[1]{\boldsymbol{#1}}		%bold-math for vectors

\begin{document}

%\title{Robust laboratory limits on a cosmological spatial gradient in the electromagnetic fine-structure constant from accelerometer experiments}

\title{\Large{
Robust laboratory limits on a cosmological spatial gradient in the electromagnetic fine-structure constant from accelerometer experiments
}}

%\author{Yevgeny~V.~Stadnik}

\author{{\large\bf \rule[30pt]{0pt}{0pt}
Yevgeny~V.~Stadnik 
}}
%\email{yevgenystadnik@gmail.com}

%\affiliation{Kavli Institute for the Physics and Mathematics of the Universe (WPI), The University of Tokyo Institutes for Advanced Study, The University of Tokyo, Kashiwa, Chiba 277-8583, Japan}

\affiliation{{\large \rule[25pt]{0pt}{0pt}
Kavli Institute for the Physics and Mathematics of the Universe (WPI), The University of Tokyo Institutes for Advanced Study, The University of Tokyo, Kashiwa, Chiba 277-8583, Japan
}}

\raggedbottom

%\date{\normalsize{
%\today
%}}

%\vspace{2mm}

%\begin{singlespace}
%\begin{abstract}
% \setlength{\leftmargin}{0mm}% <---------- CHANGE HERE
%    \setlength{\rightmargin}{\leftmargin}%
%\large
%\vspace{3mm}
%Quasar absorption spectral data indicate the presence of a spatial gradient in the electromagnetic fine-structure constant $\alpha$ on cosmological length scales. 
%We point out that experiments with accelerometers, including torsion pendula and atom interferometers, can be used as sensitive probes of cosmological spatial gradients in the fundamental constants of nature, which give rise to equivalence-principle-violating forces on test masses. 
%Using laboratory data from the E{\"o}t-Wash experiment, we constrain spatial gradients in $\alpha$ along any direction to be $| \v{\nabla} \alpha / \alpha | < 6.6 \times 10^{-4}~(\textrm{Glyr})^{-1}$ at $95\%$ confidence level. 
%Our result represents an order of magnitude improvement over laboratory bounds from clock-based searches for a spatial gradient in $\alpha$ directed along the observed cosmological $\alpha$-dipole axis. 
%\end{abstract}
%\end{singlespace}

\maketitle

 %% "Manual" Abstract %%
\vspace{3mm}
\begin{center}
\Large{\textbf{Abstract}}
\end{center}
\large

Quasar absorption spectral data indicate the presence of a spatial gradient in the electromagnetic fine-structure constant $\alpha$ on cosmological length scales. 
We point out that experiments with accelerometers, including torsion pendula and atom interferometers, can be used as sensitive probes of cosmological spatial gradients in the fundamental constants of nature, which give rise to equivalence-principle-violating forces on test masses. 
Using laboratory data from the E{\"o}t-Wash experiment, we constrain spatial gradients in $\alpha$ along any direction to be $| \v{\nabla} \alpha / \alpha | < 6.6 \times 10^{-4}~(\textrm{Glyr})^{-1}$ at $95\%$ confidence level. 
Our result represents an order of magnitude improvement over laboratory bounds from clock-based searches for a spatial gradient in $\alpha$ directed along the observed cosmological $\alpha$-dipole axis. 
Improvements to accelerometer experiments in the foreseeable future are expected to provide sufficient sensitivity to test the cosmological $\alpha$-dipole seen in astrophysical data.

\vspace{200mm}

%\tableofcontents
%\vspace{200mm}

%=================================================================================
%\section{Introduction}
%\label{Sec:Intro}

The idea of varying fundamental ``constants'' of nature dates back to the large numbers hypothesis of Dirac, who hypothesised that the gravitational constant $G$ might be proportional to the reciprocal of the age of the universe \cite{Dirac1937,Dirac1938,Dirac1974}. 
Recent studies of absorption spectra of distant quasars located in different regions of the universe \cite{Webb2011-dipole,Martins2016-dipole,Webb2020-dipole} indicate the presence of a spatial gradient in the electromagnetic fine-structure constant $\alpha = e^2 / (\hbar c)$ on cosmological length scales, where $e$ is the elementary electric charge, $\hbar$ is the reduced Planck constant and $c$ is the speed of light in vacuum. 
Variations of the fundamental constants at fixed momentum scale are not predicted within the standard model, suggesting new physics beyond the standard model. 
Varying fundamental constants are predicted in models containing new low-mass scalar particles that are also excellent candidates to explain the observed dark energy \cite{Chiba2002-quintessence,Lee2004-quintessence,Olive2008-environmental} and dark matter \cite{Menezes2005-monopoles,Stadnik2015-DM-VFCs}, which remain two of the most important outstanding problems in contemporary physics \cite{PDG2020-RPP}. 
Therefore, there is strong motivation to independently test the findings of Refs.~\cite{Webb2011-dipole,Martins2016-dipole,Webb2020-dipole} using non-astrophysical methods. 
In this paper, we consider novel non-astrophysical approaches to probe cosmological spatial gradients in the fundamental constants of nature. 
Using existing accelerometer data, we derive bounds on spatial gradients in $\alpha$ that improve over bounds from previously considered laboratory methods by an order of magnitude. 

The spatial gradient in $\alpha$ seen in Refs.~\cite{Webb2011-dipole,Martins2016-dipole,Webb2020-dipole} has a dipolar structure, a significance of $\sim 4 \sigma$ and the following magnitude: 
\begin{equation}
\label{cosmol-alpha-dipole}
\left| \frac{ \v{\nabla} \alpha }{ \alpha } \right|_\textrm{quasars}  \approx  10^{-6}~(\textrm{Glyr})^{-1}  \, . 
\end{equation}
The directional properties of this cosmological $\alpha$-dipole are summarised in Table~\ref{tab:alpha-dipole-data}. 
If the form of the $\alpha$-dipole, which is observed to be constant over a range of cosmological length scales, remains unchanged within the Solar System, then it is possible to independently test this $\alpha$-dipole using non-astrophysical methods. 
Previous ideas to test this cosmological $\alpha$-dipole using laboratory or terrestrial measurements have focused exclusively on searches for apparent temporal variations of $\alpha$ correlated with the motion of a detector along the $\alpha$-dipole axis (see Ref.~\cite{Berengut2012-dipole} and references therein). 
The drawbacks of such tests are twofold:~(i) the velocities of the Solar System and its constituent bodies are highly non-relativistic, suppressing the magnitude of the apparent temporal variations of $\alpha$ that would be seen on Earth by the factor of $v/c \ll 1$, 
and (ii) the rectilinear motion of the Sun is practically perpendicular to the axis of the cosmological $\alpha$-dipole (see Table~\ref{tab:alpha-dipole-data}), further suppressing the magnitude of the apparent temporal variations of $\alpha$ that would be seen on Earth. 
Tests via the Oklo phenomenon \cite{Shlyakhter1976-Oklo,Damour-Dyson1996-Oklo} and meteorite dating measurements \cite{Wilkinson1958-meteorites,Dyson1972-meteorites} additionally require sufficiently precise knowledge of the Solar System's trajectory over the past few billion years.

%%%%%%%%%%
\begin{table}[h!]
\centering
\caption{ \normalsize
Summary of values for the right ascension and declination of the cosmological $\alpha$-dipole in equatorial coordinates, the angle $\psi$ between the direction of increasing $\alpha$ along the $\alpha$-dipole axis and the direction of the Sun's motion, the angle $\chi$ between the direction of increasing $\alpha$ along the $\alpha$-dipole axis and the direction of the north ecliptic pole, and the approximate day on which the maximal value of $\alpha$ would be seen on Earth due to the periodic orbital motion of Earth around the Sun. 
The current direction of the Sun's motion relative to the cosmic microwave background frame has right ascension $168^\circ$ and declination $-7^\circ$ in equatorial coordinates \cite{Planck2018}. 
All of the indicated uncertainties are $1 \sigma$. 
}
\label{tab:alpha-dipole-data}
\vspace{5mm}
\large
\begin{tabular}{ |c|c|c|c|c|c| }%
\hline
 Reference & Right ascension & Declination & $\cos(\psi)$ & $\chi$ & $\alpha_\textrm{max}$ day  \\ \hline 
 \cite{Webb2011-dipole} & $17.5 \pm 0.9~\textrm{h}$ & $-58^\circ \pm 9^\circ$ & $0.06 \pm 0.13$ & $125^\circ$ & $14~\textrm{June}$  \\ \hline 
 \cite{Martins2016-dipole} & $17.2 \pm 0.7~\textrm{h}$ & $-58^\circ \pm 7^\circ$ & $0.10 \pm 0.10$ & $125^\circ$ & $11~\textrm{June}$  \\ \hline 
 \cite{Webb2020-dipole} & $16.76 \pm 1.17~\textrm{h}$ & $-63.79^\circ \pm 10.30^\circ$ & $0.16 \pm 0.12$ & $131^\circ$ & $8~\textrm{June}$  \\ \hline 
\multicolumn{1}{c}{} & \multicolumn{1}{c}{} & \multicolumn{1}{c}{} & \multicolumn{1}{c}{}  \\ 
\end{tabular}
\end{table}

The Solar System's barycentre, which coincides roughly with the position of the Sun, moves in an approximately rectilinear manner (on a laboratory timescale) relative to the comoving cosmic rest frame defined by the observed cosmic microwave background (CMB), at a speed of $370~\textrm{km/s}$ \cite{Planck2018}. 
The angle $\psi$ between the direction of increasing $\alpha$ along the $\alpha$-dipole axis and the direction of the Sun's motion has a mean value of $\cos (\psi) \sim 0.1$ (see Table~\ref{tab:alpha-dipole-data}), with the large uncertainty in $\cos (\psi)$ dominated by the uncertainty in the measured position of the cosmological $\alpha$-dipole on the sky. 
Clock-based searches for a temporal variation of $\alpha$ constrain linear-in-time drifts in $\alpha$ to $ | \dot{\alpha} / \alpha | < 4.9 \times 10^{-17}~\textrm{yr}^{-1}$ at $95\%$ confidence level \cite{NIST2008-clocks,NPL2014-clocks,PTB2014-clocks}, assuming that variations of the fundamental constants reside mainly in the electromagnetic sector. 
This translates into the following figure of merit for the sensitivity of these clock-based data to a spatial gradient in $\alpha$ directed along the cosmological $\alpha$-dipole axis: 
\begin{equation}
\label{clocks-figure-of-merit}
\left| \frac{ \v{\nabla} \alpha }{ \alpha } \right|_\textrm{clocks}  \sim  4 \times 10^{-4}~(\textrm{Glyr})^{-1}  \, . 
\end{equation}
We note that the figure of merit in Eq.~(\ref{clocks-figure-of-merit}) does not constitute a robust limit, since the assumed mean value of $\cos (\psi) \sim 0.1$ is within $\sim 1 \sigma$ of $\cos (\psi) = 0$, in which case the rectilinear motion of the Sun is perpendicular to the cosmological $\alpha$-dipole axis and the sensitivity of clock-based measurements on Earth degrades significantly. 
Furthermore, the observation of a linear-in-time drift in $\alpha$ in the laboratory by itself would not provide a confirmation of the cosmological $\alpha$-dipole seen in Refs.~\cite{Webb2011-dipole,Martins2016-dipole,Webb2020-dipole} with the current uncertainty in $\cos (\psi)$, since one would not be able to precisely infer the magnitude of the $\alpha$-dipole in this case, let alone confirm the sign of the $\alpha$-dipole. 
Likewise, the non-observation of a linear-in-time drift in $\alpha$ in the laboratory by itself would not refute the cosmological $\alpha$-dipole seen in \cite{Webb2011-dipole,Martins2016-dipole,Webb2020-dipole} with the current uncertainty in $\cos (\psi)$. 

On the other hand, one may place reasonably robust limits on a spatial gradient in $\alpha$ directed along the cosmological $\alpha$-dipole axis via laboratory searches for apparent temporal variations of $\alpha$ correlated with Earth's orbital motion around the Sun (which involves circular rather than rectilinear motion). 
The angle between the direction of increasing $\alpha$ along the $\alpha$-dipole axis and the direction of the north ecliptic pole is $\chi \approx 127^\circ$ (see Table~\ref{tab:alpha-dipole-data}), which leads to a displacement of $ \approx 2 \sin (\chi)~\textrm{AU} \approx 2.4 \times 10^{11}~\textrm{m}$ along the $\alpha$-dipole axis over the course of a year. 
Fitting the Al$^+$/Hg$^+$ clock-comparison data of Ref.~\cite{NIST2008-clocks} to the profile $\delta \alpha / \alpha = \beta \cos[ 2 \pi t / (1 \, \textrm{yr}) + \phi ]$, where the phase $\phi$ is determined by the requirement that the maximal value of $\alpha$ seen on Earth should occur on approximately 11 June (see Table~\ref{tab:alpha-dipole-data}), and assuming that the sensitivity coefficient to $\alpha$ variations is $K_{\alpha}(\textrm{Al}^+) - K_{\alpha}(\textrm{Hg}^+) \approx +3.0$ \cite{Dzuba2009-sens-coeffns}, gives $\beta = (-1.2 \pm 2.4) \times 10^{-17}~(1 \sigma)$. 
This translates into the following bound on a spatial gradient in $\alpha$ directed along the cosmological $\alpha$-dipole axis at $95\%$ confidence level: 
\begin{equation}
\label{clocks-robust-limit}
\left| \frac{ \v{\nabla} \alpha }{ \alpha } \right|_{\textrm{Al}^{+}/\textrm{Hg}^{+}} <  4.7 \times 10^{-3}~(\textrm{Glyr})^{-1}  \, . 
\end{equation}
The limit in Eq.~(\ref{clocks-robust-limit}) is an order of magnitude less stringent than the figure of merit in (\ref{clocks-figure-of-merit}) and is lacking about four orders of magnitude in sensitivity to test the cosmological $\alpha$-dipole seen in Refs.~\cite{Webb2011-dipole,Martins2016-dipole,Webb2020-dipole}, Eq.~(\ref{cosmol-alpha-dipole}). 

In this paper, we propose a different approach to robustly test the cosmological $\alpha$-dipole, Eq.~(\ref{cosmol-alpha-dipole}), in the laboratory that avoids the issues associated with the conventional tests discussed above. 
Specifically, we propose the use of accelerometers, including torsion pendula and atom interferometers, to search for the equivalence-principle-violating forces that would be exerted on two different test masses in the presence of a spatial gradient in $\alpha$. 
Searching for such equivalence-principle-violating forces with accelerometers does not rely on any motion of the apparatus along the cosmological $\alpha$-dipole axis, thereby avoiding the usual non-relativistic suppression factor $v/c \ll 1$ that is present in conventional laboratory tests. 
Additionally, the rotation of Earth (and in some experiments, the apparatus itself, e.g., via a rotating turntable) causes these equivalence-principle-violating forces to be appreciably aligned with the sensitivity axis or plane of the accelerometer for $\mathcal{O}(50\%)$ of the time, thereby avoiding the usual suppression factor associated with $\cos (\psi) \approx 0$ that plagues conventional laboratory tests. 

A number of high-precision accelerometer-based tests of the equivalence principle using Earth as the attractor have been performed, including the laboratory-based E{\"o}t-Wash experiment \cite{Eot-Wash2008-TP}, the space-based MICROSCOPE mission \cite{MICROSCOPE2017-TP}, and the recent atom-interferometry measurements in the laboratory reported in \cite{Stanford2020-AI}. 
To illustrate the basic principles of accelerometer-based searches for cosmological spatial gradients in $\alpha$, we focus on the E{\"o}t-Wash measurements reported in Ref.~\cite{Eot-Wash2008-TP}, which have the best sensitivity to differential accelerations along the sensitivity plane of the apparatus and lead to the most stringent accelerometer-based bounds on cosmological spatial gradients in $\alpha$. 
Our main result is summarised in Table~\ref{tab:results_summary}, along with the sensitivity estimates for other accelerometer-based experiments with existing datasets.

Henceforth, we shall consider variations of the fundamental constants in the non-relativistic limit and, unless explicitly stated otherwise, we shall adopt the natural system of units $\hbar = c = 1$. 
A test particle or test body of mass $M$, which varies in space or time, experiences the following additional acceleration in the non-relativistic limit (see, e.g., Refs.~\cite{Damour-1996-EP,Uzan-2011-review} and references therein): 
\begin{equation}
\label{varying-mass-acceleration}
\delta \v{a} = - \frac{\v{\nabla} M}{M} - \frac{\dot{M}}{M} \v{v}  \, , 
\end{equation}
where $\v{v}$ is the velocity of the test particle or body with respect to the comoving cosmic rest frame, which we again take to be the CMB frame. 
The physical meaning of the first term in Eq.~(\ref{varying-mass-acceleration}) is that a test particle is attracted towards the direction where the particle has a lower mass-energy, with the particle's mass energy being converted into kinetic energy in the process, while the second term in (\ref{varying-mass-acceleration}) follows from conservation of linear momentum. 
The mass-energy of a non-relativistic electrically-neutral atom containing $A$ nucleons and $Z \gg 1$ electrons can be approximated as: 
\begin{equation}
\label{atom-mass-formula}
M_\textrm{atom} \approx A m_N + Z m_e + \frac{a_C Z^2}{A^{1/3}}  \, , 
\end{equation}
where we have neglected smaller electromagnetic mass-energy contributions, such as the electronic binding energy and the electromagnetic energies of the individual nucleons. 
The first two terms in Eq.~(\ref{atom-mass-formula}) correspond to the nucleon and electron rest-mass-energies, $m_N$ and $m_e$, respectively. 
The third term in (\ref{atom-mass-formula}) corresponds to the energy associated with the electrostatic repulsion between protons in a spherical nucleus of uniform electric-charge density, with the coefficient $a_C \approx 3 \alpha / (5 r_0) \approx 0.7~\textrm{MeV}$, where $r_0 \approx 1.2~\textrm{fm}$ is the internucleon separation parameter that is determined chiefly by the strong nuclear force. 

Since the fractional mass-energy contributions due to the electromagnetic, electron-mass and nucleon-mass components in Eq.~(\ref{atom-mass-formula}) generally differ for different test particles or test bodies, different particles or bodies will therefore experience different accelerations via Eq.~(\ref{varying-mass-acceleration}). 
The equivalence-principle-violating forces resulting from a cosmological spatial gradient in one or more of the fundamental constants of nature can be sought with accelerometers employing two different test-particle species or two bodies of different material compositions. 
In order to circumvent possible degeneracies associated with accidental cancellations between variations of $\alpha$ and other fundamental constant(s) for a single test-mass pair, one can repeat measurements using different test-mass pairs. 
At the time of writing, there does not appear to be strong evidence of variations of the fundamental constant(s) other than $\alpha$. 
Therefore, if variations of the fundamental constants reside predominantly in the electromagnetic sector, 
then the difference in acceleration between two test particles or test bodies reads as follows:\footnote{\normalsize We remark that the derivation of the non-relativistic result (\ref{differential-acceleration-formula-MASTER}) in the limit of zero momentum does not require the specification of the model or Lagrangian that sources the $\alpha$ variation. 
On the other hand, relativistic corrections to Eq.~(\ref{differential-acceleration-formula-MASTER}) that arise at non-zero values of momentum do require such a specification, due to possible modifications to the form of electrodynamics for relativistic fermions (see, e.g., Ref.~\cite{Bekenstein2003-magic}) and changes to the form of the running of $\alpha$ with momentum scale. } 
\begin{equation}
\label{differential-acceleration-formula-MASTER}
\delta (\v{a}_1 - \v{a}_2) \approx \frac{ \left[ (A_1 W_2 - A_2 W_1) m_N + (Z_1 W_2 - Z_2 W_1) m_e \right] a_C }{ (A_1 m_N + Z_1 m_e + W_1 a_C) (A_2 m_N + Z_2 m_e + W_2 a_C) } \left( \frac{ \v{\nabla} \alpha }{ \alpha } + \frac{ \dot{\alpha} }{ \alpha } \v{v} \right)  \, , 
\end{equation}
where $W_i = Z_i^2 / A_i^{1/3}$. 
In the case of the E{\"o}t-Wash measurements in \cite{Eot-Wash2008-TP}, which employed beryllium and titanium test bodies, Eq.~(\ref{differential-acceleration-formula-MASTER}) reads: 
\begin{equation}
\label{differential-acceleration-formula-Be-Ti}
\delta (\v{a}_\textrm{Be} - \v{a}_\textrm{Ti}) \approx + 1.4 \times 10^{-3}  \left( \frac{ \v{\nabla} \alpha }{ \alpha } + \frac{ \dot{\alpha} }{ \alpha } \v{v} \right)  \, . 
\end{equation}

The torsion-pendulum measurements in \cite{Eot-Wash2008-TP} constrained space-fixed differential accelerations in any direction to $ | \delta (\v{a}_\textrm{Be} - \v{a}_\textrm{Ti}) | < 8.8 \times 10^{-15}~\textrm{m/s}^2$ at $95\%$ confidence level. 
In the limiting case that variations of $\alpha$ are purely spatial in the comoving cosmic rest frame, then using Eq.~(\ref{differential-acceleration-formula-Be-Ti}), we derive the following robust limit on a spatial gradient in $\alpha$ \textit{along any direction} at $95\%$ confidence level:\footnote{\normalsize Motion of the apparatus along a spatial gradient in $\alpha$ can give rise to temporal changes in the apparent size of the spatial gradient in $\alpha$ between two observers using different references for the unit of length (or equivalently the unit of time, if $c$ remains constant). In the non-relativistic limit, the length of a solid object scales as $\propto 1/(m_e \alpha)$, while lengths defined via an optical or hyperfine atomic transition frequency scale as $\propto 1/(m_e \alpha^2)$ and $\propto m_p/(m_e^2 \alpha^4)$, respectively. Apparent linear-in-time drifts in $\alpha$ and $m_e/m_p$ over the duration of the E{\"o}t-Wash measurements are independently constrained to be very small, making any such reference-dependent changes to the numerical value quoted in Eq.~(\ref{Eot-Wash-robust-limit}) negligible. } 
\begin{equation}
\label{Eot-Wash-robust-limit}
\left| \frac{ \v{\nabla} \alpha }{ \alpha } \right|_{\textrm{Be}-\textrm{Ti}} <  6.6 \times 10^{-4}~(\textrm{Glyr})^{-1}  \, . 
\end{equation}
The limit in Eq.~(\ref{Eot-Wash-robust-limit}) is an order of magnitude more stringent than the clock-based limit in Eq.~(\ref{clocks-robust-limit}), which applies to a spatial gradient in $\alpha$ directed along the axis of the cosmological $\alpha$-dipole observed in Refs.~\cite{Webb2011-dipole,Martins2016-dipole,Webb2020-dipole}. 
Since the bound (\ref{Eot-Wash-robust-limit}) applies to spatial gradients in $\alpha$ along any direction, it should be regarded as a conservative limit on a spatial gradient in $\alpha$ directed along the axis of the cosmological $\alpha$-dipole seen in \cite{Webb2011-dipole,Martins2016-dipole,Webb2020-dipole}; a separate analysis specifically for the axis direction of the cosmological $\alpha$-dipole observed in \cite{Webb2011-dipole,Martins2016-dipole,Webb2020-dipole} may give a more stringent torsion-pendulum-based limit than in Eq.~(\ref{Eot-Wash-robust-limit}).

%%%%%%%%%%
\begin{table}[h!]
\centering
\caption{ \normalsize
Summary of bounds on a cosmological $\alpha$-dipole from terrestrial experiments and analogous sensitivity estimates based on existing datasets. 
All of the bounds and sensitivity estimates are at the $95\%$ confidence level. 
The $\alpha$-variation sensitivity coefficients for the accelerometer-based experiments are defined by the dimensionless prefactor that appears in the right-hand side of Eq.~(\ref{differential-acceleration-formula-MASTER}). 
The figures for accelerometer-based experiments apply to spatial gradients in $\alpha$ along an arbitrary space-fixed direction, while the atomic-clock-based bound assumes a spatial gradient in $\alpha$ directed along the cosmological $\alpha$-dipole seen in Refs.~\cite{Webb2011-dipole,Martins2016-dipole,Webb2020-dipole}. 
The partial dataset of MICROSCOPE reported in Ref.~\cite{MICROSCOPE2017-TP} constitutes $\approx 7 \%$ of the total dataset acquired during the whole mission \cite{MICROSCOPE2019-TP}. 
}
\label{tab:results_summary}
\vspace{5mm}
\large
\begin{tabular}{ |c|c|c| }%
\hline
 Experiment & Sensitivity coefficient & Bound on $\left| \v{\nabla} \alpha / \alpha \right| / (\textrm{Glyr})^{-1}$  \\ \hline 
 E{\"o}t-Wash (Be-Ti) & $+1.4 \times 10^{-3}$ & $6.6 \times 10^{-4}$   \\ \hline 
 Atomic clocks (Al$^+$/Hg$^+$) & --- & $4.7 \times 10^{-3}$   \\ \hline 
 MICROSCOPE (Ti-Pt, partial dataset) & $+1.9 \times 10^{-3}$ & $\sim 2 \times 10^{-2}$ (estimate)   \\ \hline 
 MICROSCOPE (Ti-Pt, full dataset) & $+1.9 \times 10^{-3}$ & $\sim 6 \times 10^{-3}$ (estimate)   \\ \hline 
 Atom interferometry ($^{85}$Rb-$^{87}$Rb) & $-8 \times 10^{-5}$ & $\sim 2 \times 10^2$ (estimate)   \\ \hline 
\multicolumn{1}{c}{} & \multicolumn{1}{c}{} & \multicolumn{1}{c}{}   \\ 
\end{tabular}
\end{table}

Let us briefly explain why the E{\"o}t-Wash measurements in \cite{Eot-Wash2008-TP} lead to the most stringent accelerometer-based limit on a cosmological spatial gradient in $\alpha$, even though the MICROSCOPE mission \cite{MICROSCOPE2017-TP} gives a more stringent limit on equivalence-principle-violating forces that are \textit{directed radially} towards Earth's center. 
The combined rotation of Earth and the apparatus in the E{\"o}t-Wash experiment causes a cosmological spatial gradient in $\alpha$ \textit{along any space-fixed direction} to be appreciably aligned with the horizontal sensitivity plane of the apparatus (which is tilted by $\sim 10^{-3}~\textrm{rad}$ away from the vertical direction due to Earth's rotation) for $\mathcal{O}(50\%)$ of the time, thereby avoiding the $\sim 10^{-3}$ suppression factor that is present in tests of the equivalence principle that use Earth as the attractor and search for radially-directed forces (this suppression factor is absent in atom-interferometry experiments\footnote{\normalsize In laboratory-based atom-interferometry experiments, the sensitivity axis of the apparatus is oriented vertically and there is no rotation of the apparatus in the laboratory frame of reference.} and in the space-based MICROSCOPE experiment). 
Since there are no such suppression factors in the E{\"o}t-Wash experiment, MICROSCOPE mission or atom-interferometry experiments when searching for a cosmological spatial gradient in $\alpha$ directed along a \textit{space-fixed} direction, the E{\"o}t-Wash experiment gains a factor of $\sim 10^3$ in sensitivity compared with the latter types of experiments, when searching for a cosmological spatial gradient in $\alpha$ instead of \textit{radially-directed} equivalence-principle-violating forces.

In terms of possible improvements, pairing a test mass consisting of very-low-$Z$ element(s) with a test mass consisting of very-high-$Z$ element(s), such as in the Be-Pt test-mass pair, can provide an increase over the $\alpha$-variation sensitivity coefficient for the Be-Ti test-mass pair in Eq.~(\ref{differential-acceleration-formula-Be-Ti}) by up to a factor of a few. 
By implementing a combination of upgrades that have been demonstrated experimentally but have not yet been specifically implemented in torsion-pendulum experiments, the sensitivity of torsion-pendulum experiments may be improved by about three orders of magnitude compared to our bound in (\ref{Eot-Wash-robust-limit}), which would allow torsion-pendulum experiments to directly test the cosmological $\alpha$-dipole observed in Refs.~\cite{Webb2011-dipole,Martins2016-dipole,Webb2020-dipole}, Eq.~(\ref{cosmol-alpha-dipole}). 
These upgrades can be realised in ground-based experiments and involve upgrades to the angle read-out system, increases in the quality factor and turntable frequency of the torsion pendulum, as well as the use of different materials for the suspension fibres; see Refs.~\cite{Wagner2012-TP-review,Arp2013-TP-upgrade,Graham2016-DM-accelerometers} for further details. 
A further improvement in sensitivity to $\v{\nabla} \alpha / \alpha$ by up to another three orders of magnitude may be achieved via further improvements to the angle read-out system and increases in the quality factor and turntable frequency of the torsion pendulum, as well as the use of more massive test masses (along with a correspondingly larger torsional spring constant) and cryogenic cooling of the system from room temperature; see Refs.~\cite{Graham2016-DM-accelerometers,Hogan2011-TP-upgrade} for more details. 
Cryogenic cooling can also be implemented in space-based experiments, such as the STEP mission~\cite{STEP2001-EP,STEP2007-EP}, which together with the removal of the ground wire for the proof masses (which has recently been demonstrated by the LISA Pathfinder mission~\cite{LISA-Pathfinder2018}) would allow for an improvement in sensitivity by three orders of magnitude compared to the MICROSCOPE mission.

Meanwhile, the sensitivity of interferometry experiments involving two different species of cold atoms has improved by five orders of magnitude over the past several years alone \cite{Rasel2014-AI,Tino2014-AI,Wuhan2015-AI,Tino2017-AI,Stanford2020-AI} and is now approaching that of torsion-pendulum tests of the equivalence principle \cite{Eot-Wash2008-TP,MICROSCOPE2017-TP,Wagner2012-TP-review}. 
It appears challenging for ground-based atom interferometers to achieve sufficient sensitivity to test the cosmological $\alpha$-dipole observed in Refs.~\cite{Webb2011-dipole,Martins2016-dipole,Webb2020-dipole}. 
However, a number of space-based atom-interferometric missions (including AEDGE \cite{AEDGE2020-AI}, AGIS \cite{AGIS2008-AI}, AIGSO \cite{AIGSO2018-AI} and SAGE \cite{SAGE2019-AI}), which have been proposed to search for gravitational waves, would exceed the sensitivity required to test the cosmological $\alpha$-dipole in Eq.~(\ref{cosmol-alpha-dipole}) by at least a few orders of magnitude.

Finally, in the limiting case that variations of $\alpha$ are purely temporal in the CMB frame, then using Eq.~(\ref{differential-acceleration-formula-Be-Ti}) and noting that the speed of the Solar System with respect to the CMB frame is $370~\textrm{km/s}$ \cite{Planck2018}, we derive the following limit on a linear-in-time drift in $\alpha$ at $95\%$ confidence level using the torsion-pendulum data in \cite{Eot-Wash2008-TP}: 
\begin{equation}
\label{Eot-Wash-limit-temporal}
\left| \frac{ \dot{\alpha} }{ \alpha } \right|_{\textrm{Be}-\textrm{Ti}} <  5.4 \times 10^{-10}~\textrm{yr}^{-1}  \, . 
\end{equation}
The limit in Eq.~(\ref{Eot-Wash-limit-temporal}) is several orders of magnitude less stringent than bounds from clock-comparison measurements \cite{NIST2008-clocks,NPL2014-clocks,PTB2014-clocks}, due to the $\sim 10^{-3}$ suppressed sensitivity coefficient and the additional non-relativistic $v/c \sim 10^{-3}$ suppression factor in Eq.~(\ref{differential-acceleration-formula-Be-Ti}), the former of which arises because the electromagnetic mass-energy contribution to the overall mass of an atom is small to begin with, see Eq.~(\ref{atom-mass-formula}).

%%%%%%%%%%%%%%%%%%%%%%
%\section*{Note}
\textit{Note added} --- 
Most recently, two new relevant sets of clock-based data were brought to my attention \cite{BACON2021-clocks,PTB2021-clocks}. 
These new data may also be analysed to search for a cosmological spatial gradient in $\alpha$.

%%%%%%%%%%%%%%%%%%%%%%
%\section*{Acknowledgements}
\textbf{Acknowledgements} --- 
I thank Victor Flambaum for helpful discussions. 
I thank David Hume for providing additional information related to the Al$^+$/Hg$^+$ spectroscopy data in Ref.~\cite{NIST2008-clocks}. 
I thank Erik Shaw for discussions about the current status of the E{\"o}t-Wash experiments. 
This work was supported by the World Premier International Research Center Initiative (WPI), MEXT, Japan, and by the JSPS KAKENHI Grant Number JP20K14460.

%===================================================================================

\end{document}